%% file: main.tex
\documentclass[]{spie}  

\usepackage{amsmath,amsfonts,amssymb}
\usepackage{graphicx}
\usepackage[colorlinks=true, allcolors=blue]{hyperref}
\usepackage{booktabs}

\usepackage{silence}
\usepackage[subpreambles=false, mode=buildnew]{standalone}
\usepackage{siunitx}
\usepackage{svg}
\usepackage{caption}
\usepackage{subcaption}
\usepackage{orcidlink}
\usepackage{pgfplots}
\pgfplotsset{compat=1.14}
\usepackage{microtype}

\title{Fabry–Pérot open resonant cavities for measuring the dielectric parameters of mm-wave optical materials
}

\author[a,b]{Brodi D. Elwood\,\orcidlink{0000-0003-4117-6822}}
\author[a]{Paul K. Grimes\,\orcidlink{0000-0001-9292-6297}}
\author[a,b]{John Kovac}
\author[a]{Miranda Eiben}
\author[a,b]{Grant Meiners}
\affil[a]{Center for Astrophysics $\vert$ Harvard \& Smithsonian, Cambridge, MA 02138, USA}
\affil[b]{Department of Physics, Harvard University, Cambridge, MA 02138, USA}

\authorinfo{Send correspondence to B. D. Elwood: \href{mailto:bdelwood@fas.harvard.edu}{bdelwood@fas.harvard.edu}}

\begin{document}
\maketitle

\input{sec/abstract}
\input{sec/intro}
\input{sec/cavities}

\input{sec/measurement}
\input{sec/results}
\input{sec/conclusion}

\bibliography{main} 
\bibliographystyle{spiebib} 

\end{document}

%% file: sec/abstract.tex
\begin{abstract}
    As millimeter-wave cosmology experiments refine their optical chains, precisely characterizing their optical materials under cryogenic conditions becomes increasingly important. For instance, as the aperture sizes and bandwidths of millimeter-wave receivers increase, the design of antireflection coatings becomes progressively more constrained by an accurate measure of material optical properties in order to achieve forecasted performance. Likewise, understanding dielectric and scattering losses is relevant to photon noise modeling in presently-deploying receivers such as BICEP Array and especially to future experiments such as CMB-S4. Additionally, the design of refractive elements such as lenses necessitates an accurate measure of the refractive index. High quality factor Fabry–Pérot open resonant cavities provide an elegant means for measuring these optical properties. Employing a hemispherical resonator that is compatible with a quick-turnaround 4 Kelvin cryostat, we can measure the dielectric and scattering losses of low-loss materials at both ambient and cryogenic temperatures. We review the design, characterization, and metrological applications of quasioptical cavities commissioned for measuring the dielectric materials in the BICEP3 (95 GHz) and BICEP Array mid-frequency (150 GHz) optics. We also discuss the efforts to improve the finesse of said cavities, for better resolution of degenerate higher order modes, which can provide stronger constraints on cavity parameters and sample material thickness.
\end{abstract}

\keywords{Resonant cavity, Dielectric losses, Quasioptical, Fabry–Pérot, Optical materials, Sub-millimeter instrumentation, Complex permittivity}

%% file: sec/intro.tex
\section{INTRODUCTION}
\label{sec:intro}

The astrophysical observations carried out in the millimeter and sub-millimeter wavelength regime provide key insights into many of the open questions in physics today. Experiments using millimeter light image black holes, map high-redshift regions of the Universe, and observe the Cosmic Microwave Background, to name a few \cite{johnsonKeyScienceGoals2023,karkareSPTSLIMLineIntensity2022,bicep/keckcollaborationImprovedConstraintsPrimordial2021}. These experiments aim to expand our understanding of the influence of black holes on cosmic evolution, explore the growth of structure in the cosmos, and provide insights into the earliest epochs of the Universe. Optics used in experiments within the mm/sub-mm regime often incorporate polymers like polyethylene, polytetrafluoroethylene, and nylon; ceramics such as alumina; or monocrystalline silicon. Plastics are frequently used for filters, lenses, vacuum windows, or laminate antireflection coatings, while alumina and silicon are common filter and lens materials \cite{choiRadiotransparentMultilayerInsulation2013,dierickxPlasticLaminateAntireflective2023,barkatsUltrathinLargeapertureVacuum2018}. An experimenter's choice of a material for a given optic hinges on its complex permittivity (namely index of refraction and dielectric losses) at the desired temperature, which is often cryogenic\cite{carterLowlossSiliconOptical2024,sobrinDesignIntegratedPerformance2022,adeBicepKeckXV2022}. As mm-wave experiments often employ cryogenically cooled optics in order to reduce optical loading, we seek a robust method of empirically measuring complex permittivity as a function of temperature, as its behavior can be nontrivial \cite{paineProcessingCalibrationSubmillimeter2013}. In the context of current and future CMB experiments, such as BICEP/Keck or CMB-S4, we are motivated to seek high precision measurements of  refractive index as we require an accuracy of $\pm.005$ in index measurements for designing optics. Additionally, we desire to understand material loss in order to better model instrument photon noise and systematics, such as forebaffle coupling and far-sidelobe response.

Numerous well-established techniques exist for measuring the the index and loss of materials in the mm-wave regime. Many of those techniques utilize closed resonant cavities, typically waveguides \cite{weirAutomaticMeasurementComplex1974}. The accuracy of index and loss measurements from waveguide cavity (so-called Nicholson-Ross-Weir) methods relies on the quality of sample machining; gaps between the material under test and waveguide walls result in significant errors, which are exacerbated at high frequencies. When one wishes to measure complex permittivity at cryogenic temperatures, in the case of a closed resonator, one has to take care to account for differential coefficients of thermal expansion between the metallic waveguide and the dielectric sample. 

Additionally, due to the frequency scaling of the quality factor with the resonant frequency, and since the accuracy of loss measurements deteriorates for losses below the reciprocal of the quality factor, NRW methods become impracticable at high frequencies, typically above $\SI{75}{GHz}$. Furthermore, for low loss materials of thicknesses that are integer multiples of wavelength, measurements using NRW methods become unstable \cite{baker-jarvisImprovedTechniqueDetermining1990}. Similarly, measurement techniques employing Fourier transform spectrometers suffer from increased systematic uncertainty in the limit of low loss materials or samples thin compared to a wavelength \cite{chamberlainDeterminationRefractiveIndex1969,nadolskiBroadbandMillimeterwaveAntireflection2020}.

The use of an open resonator comes with the advantage that the standing modes of optical resonators, in particular simple two-mirror topologies such as Fabry–Pérot hemispherical open resonators, are well-studied and have readily recordable observables that depend strongly on cavity electrical length and round-trip loss \cite{cullenAccurateMeasurementPermittivity1997}. Compared to closed resonator techniques, the mode spectrum of cavities enables quasi-broadband measurements. Due to the build-up of the electromagnetic field in the cavity, we can naturally integrate over many transits through the inserted optical material of interest. High quality factor Fabry–Pérot open cavities provide a means for measuring the optical properties of low-loss dielectric materials which only relies on one length dimension of the material under test. By taking high signal-to-noise measurements of the resonant modes of an empty and dielectric-loaded open resonator, we can infer the index and loss of an inserted dielectric sample. Measurement apparatuses can be readily designed for operation down to cryogenic temperatures.

In this work, we first establish the basic working theory of quasioptical open resonant cavities, employing the full vector theory as established in the literature \cite{yuMeasurementPermittivityMeans1982}. We discuss a quasioptical cavity fabricated for this work and detail a measurement procedure for inferring index and loss from measurements of its fundamental modes, before reporting the results for several polyethylene samples. Finally, we conclude with a brief discussion of future improvements to open resonator techniques for measuring complex permittivity, including ongoing efforts to extend these methods to cryogenic temperatures.

%% file: sec/cavities.tex
\section{(Quasi)-optical cavities}
\label{sec:cavity_theory}

Given the challenges associated with traditional methods for measuring complex permittivity in the mm-wave regime, open resonant cavities offer a promising alternative due to their high-quality factors and ability to integrate over multiple transits of the optical material. To fully leverage these advantages, careful consideration must be given to the design of the cavity, particularly in minimizing power losses and ensuring effective coupling. In the end, we construct cavity quality factor such that we can measure low-loss materials while retaining enough transmitted power to stay above the transmission background of the cavity. To construct a cavity with a given quality factor, we must account for the sources of power loss intrinsic to the cavity: coupling losses from the apertures, conductive losses on the mirror surfaces, and diffractive losses off the edges of the mirror and apertures.

\begin{figure}[!ht]
    \centering
    \begin{subfigure}[t]{0.45\textwidth}
        \centering
        \scalebox{-1}[1]{\includegraphics[width=\textwidth]{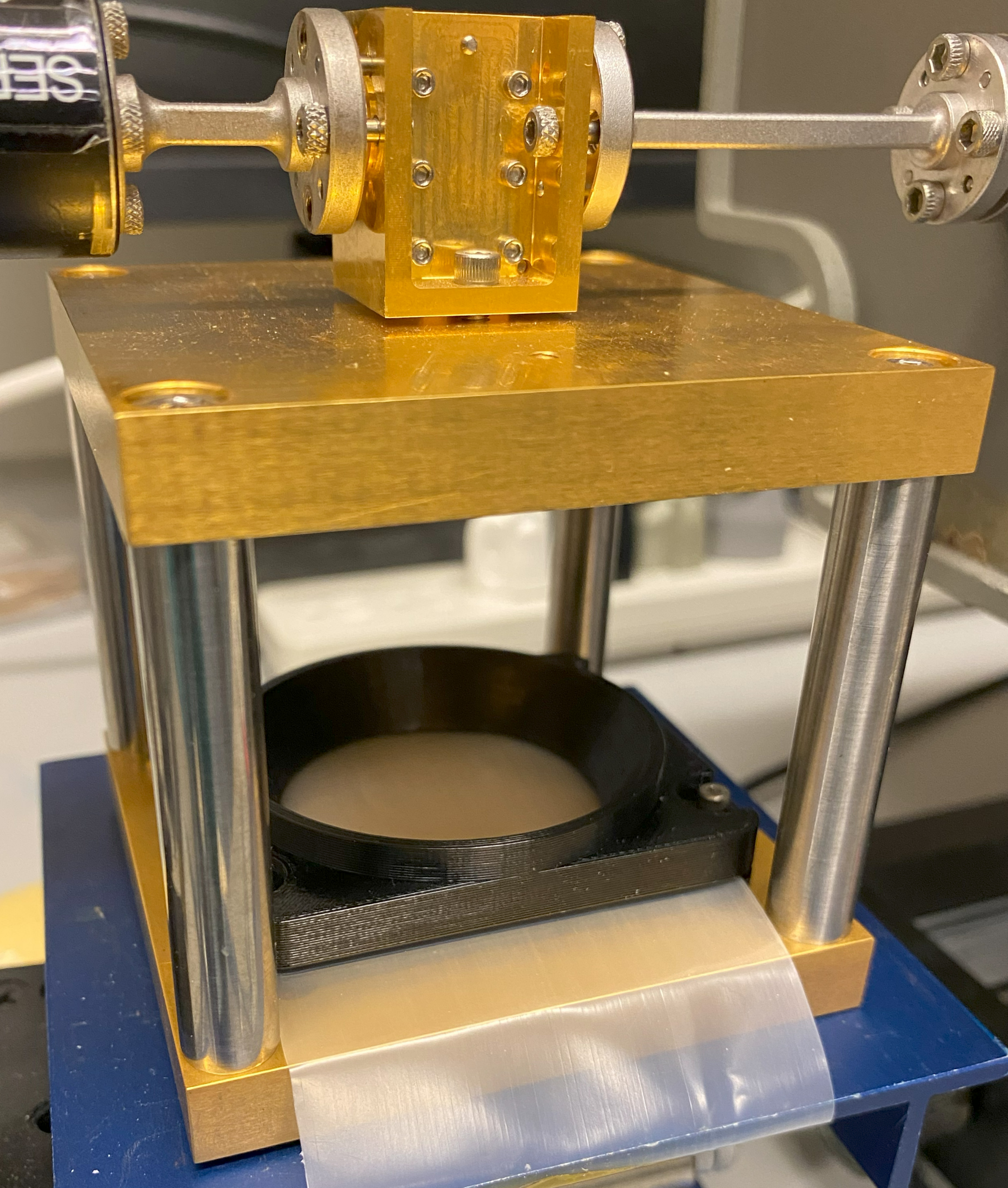}}
        \caption{Resonant cavity loaded with a thin film of LDPE.} \label{fig:cavity_photo}
    \end{subfigure}
    \hfill
    \begin{subfigure}[t]{0.5\textwidth}
        \centering
        \includestandalone[width=1.0\textwidth]{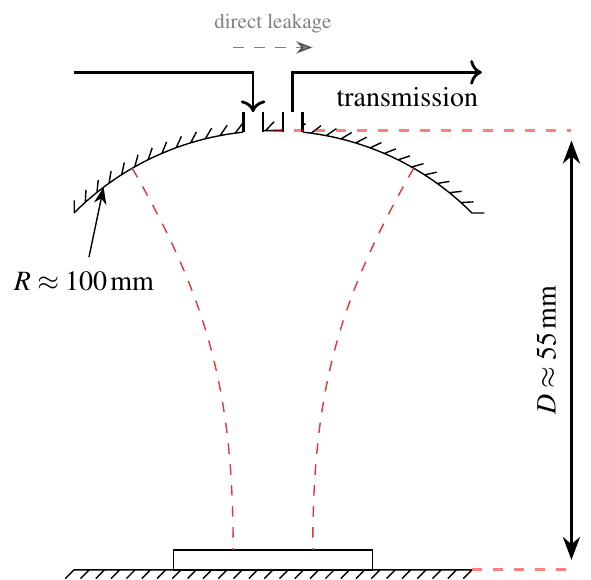}
        \caption{A pictorial representation of the FP resonator.} \label{fig:schematic}
    \end{subfigure}
    \vspace{0.5cm} 
    \begin{subfigure}[t]{\textwidth}
        \centering
        \includestandalone[width=1.0\textwidth]{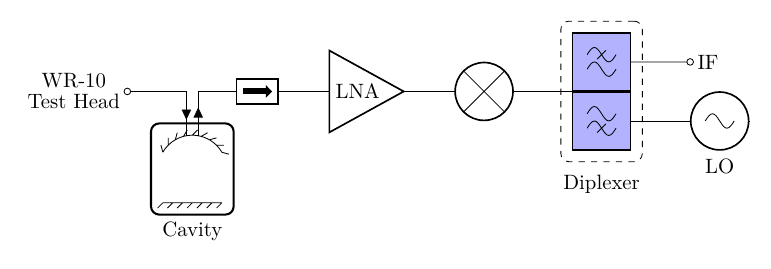}
        \caption{A block diagram of the signal path.} \label{fig:block_diagram}
    \end{subfigure}
    \caption{Gold-plated planar and spherical mirrors, separated by Invar struts, form a hemispherical open resonator. A low CTE material was chosen for the struts in order to minimize temperature-induced fluctuations in cavity length, which directly lead to fluctuations in resonant frequency. A WR-10 VNA extension head couples radiation into the cavity via a split-block waveguide butted into a recess in the top (spherical) mirror. Two circular coupling apertures are sized to set the intrinsic cavity quality factor to \num{1e5}. The cavity depicted here has a length of \SI{55}{\mm} and a mirror radius of curvature of \SI{100}{\mm}, putting the spacing between fundamental modes at $\approx$\SI{2.7}{\GHz}. Compared to a Gaussian spot size at the spherical mirror of $\SI{11.85}{\mm}$, the spherical mirror has a radius of $\SI{50}{\mm}$, minimizing diffractive losses off that surface. Dielectric samples are pressed against the planar mirror with a bracket fastened to the edges of the planar surface.}
    \label{fig:combined}
\end{figure}

The coupling constant $\beta$---the ratio of cavity incident power to intrinsic power lost---parametrizes the power coupled into the cavity. For the cavity used in this work, we selected an aperture coupling of $\beta=\num{0.3}$. Ansys HFSS calculations were carried out to show apertures of $\SI{1}{\mm}$ diameter and $\SI{1.05}{\mm}$ depth provide the expected aperture coupling. The aperture coupling dominates the quality factor of the cavity. Fig. \ref{fig:combined} gives further details on the cavity employed in this work. Practically, we are constrained to the current insertion loss due to leakage between the input and transmit sides of the split-block waveguide used to inject power into and out of the cavity: power transits directly between the input and output waveguides, bypassing the cavity entirely. This direct leakage constitutes the primary component of the noise floor of our measurements. Thus, raising the quality factor of the cavity, which necessitates transmitting less power through the cavity, also results in reducing the signal to noise ratio of the experiment. To make further increases to the quality factor useful, we must decrease direct leakage across the split-block waveguide. See Sec. \ref{sec:conclusion} for a brief discussion on future work in this area.

Light is coupled into the cavity from the vector network analyzer (VNA) extension head via a $\textrm{TE}_{10}$ mode rectangular split-block waveguide butted against circular apertures, which magnetically couples as a magnetic dipole radiator. The transverse component of the $\textrm{TE}_{10}$ mode's magnetic field excite the eigenmodes, $\textrm{TEM}_{plq}$, of the cavity, where $p$, $l$, and $q$ are the radial, azimuthal, and axial (longitudinal), mode numbers, respectively. The resonance condition for the fundamental modes $\textrm{TEM}_{00q}$ of the empty resonator is \cite{yuMeasurementPermittivityMeans1982}
\begin{align}
    \label{eq:empty_res}
    f_{q} = \frac{c}{2 D}\left(q+1+\frac{1}{\pi}\arctan{\sqrt{\frac{D}{R-D}}}-\frac{1}{2\pi k_q R}\right),
\end{align}
where $D$ is the cavity length, $R$ the spherical mirror radius of curvature, and $k$ the freespace wavenumber. When the cavity is loaded with a dielectric sample, we perturb the mode spectrum and cavity loss. As a result, the resonance condition becomes \cite{yuMeasurementPermittivityMeans1982}
\begin{align}
    \label{eqn:loaded_res}
    \frac{1}{n}\tan{(nkt-\Phi_T)=-\tan{(kd-\Phi_D)}},
\end{align}
with
\begin{align}
    \Phi_T & = \arctan{\left(\frac{t}{nz_0}\right)}  - \arctan{\left(\frac{1}{nkR_1(t)}\right)},                                     \\
    \Phi_D & = \arctan{\left(\frac{d'}{z_0}\right)} - \arctan{\left(\frac{t}{n z_0}\right)} - \arctan{\left(\frac{1}{k R}\right)}  +
    \arctan{\left(\frac{1}{k R_2(t)}\right)},                                                                                        \\
    R_1(t) & = t + \frac{n^2 z_0^2}{t},                                                                                              \\
    R_2(t) & = \frac{R_1(t)}{n},                                                                                                     \\
    z_0    & = \sqrt{d'(R-d')},                                                                                                      \\
    d'     & = d + \frac{t}{n},
\end{align}
and
\begin{align}
    d & = D-t.
\end{align}
From the above, one can see that $\Phi_T$ and $\Phi_D$ encode information on cavity geometry and the dielectric electrical length (through index $n$ and sample thickness $t$).

The solutions of Eqn. \eqref{eqn:loaded_res}, $f_{L,i}$, give an initial estimate of the resonant frequencies of the fundamental
modes of the loaded cavity. This equation arises from matching the cavity field configurations at the free
space-dielectric interface and at the mirror boundary on the axis of the resonator. This estimate assumes the surface of the dielectric matches the wavefront exactly, and that the wavefront has no tangential component at the spherical mirror boundary. With a flat sample, there is some mismatch between the Gaussian wavefront and the medium interface. Likewise, since the Gaussian wavefront does not quite match the spherical mirror surface, it has some tangential component at the mirror boundary. Both of these effects result in a shift of the loaded resonant frequencies. We can correct the initial estimate as

\begin{align}
    f_{L} & = f_{L,i} (1 + f_{\textrm{interface}} + f_{\textrm{mirror}})                                               \\
    f_{L} & = f_{L,i}\left(1 + \frac{t(n-\Delta)}{n^2k^2 \omega_t^2(t\Delta+d)} + \frac{3}{4k^2(t\Delta + d)R}\right),
\end{align}
where the beam radius at the dielectric interface is
\begin{align}
    \omega_t = w_0 \sqrt{1 + \left(\frac{t}{z_0}\right)^2},
\end{align}
and
\begin{align}
    \Delta\equiv\frac{n^2}{n^2 \cos^2(nkt-\Phi_T)+\sin^2(nkt-\Phi_T)}.
\end{align}
Finally, we compute the loss tangent as
\begin{align}
    \label{eq:losstan}
    \tan\delta = \frac{1}{Q_s}\frac{2nk(t\Delta+d)}{2nkt\Delta-\Delta\left[\sin{2(nkt-\Phi_T)}\right]},
\end{align}
where $\frac{1}{Q_s}=\frac{1}{Q_{L}}-\frac{1}{Q_0(k)}$ \cite{hirvonenMeasurementDielectrics1001996}. In principle, the intrinsic quality factor, $Q_0(k)$, can be computed by modeling and simulating the cavity in electromagnetic simulation software. However, the intrinsic quality factor is dominated by the evanescent coupling through the apertures in the hemispherical mirror, which is extremely sensitive to fabrication tolerances. In practice, it is sufficient to interpolate the $Q_0(k)$ curve using measurements of adjacent fundamental modes of the cavity.

%% file: sec/measurement.tex
\section{Measurement procedure}

We excite the cavity through the coupling block with a WR-10 VNA extension head, and measure the cavity response using a custom-built receive-only VNA extension head, both attached to an HP 8510C millimeter-wave VNA. There are two measurement procedures for sweeping the cavity response---keeping the cavity length fixed while sweeping in frequency, or adjusting cavity length while injecting a single tone. The former procedure has the advantage of avoiding the complications of precise length modulation, while the latter has the advantage of maintaining the intrinsic quality factor, $Q_0(k)$, due to the identical aperture coupling for the loaded and unloaded cavity resonances, as the frequencies of the resonances remain static. In this work, we adopt the frequency-variation method, and interpolate the intrinsic quality factor measured  to the shifted resonances.

Using a VNA for carrying out these measurements comes with several other advantages. For one, modern systems carry out averaging, which can suppress readout noise. With high quality factor cavities, this is important as only a small amount of power is leaked from the cavity. More importantly, we can capture, with a high precision in frequency, the cavity response, which allows us to employ fitting techniques on the cavity mode spectrum in order to infer cavity and mode parameters such as cavity length, mirror radius of curvature, and quality factor.

The resonance modes of quasioptical cavities can be well-modeled by a so-called Fano line\cite{petersanMeasurementResonantFrequency1998},

\begin{align}
  S_t(f)=\frac{S_M}{1+2iQ\frac{f-f_0}{f_0}}+S_c,
\end{align}
where $S_t$ is the measurement scattering coefficient $S_{12}$ or $S_{21}$. We can fit to the amplitude and the phase. $f_0$ and $Q$ are the mode frequency and quality factor, respectively. The Fano lineshape is nearly a Lorentzian, but admits a complex coupling factor, $S_c$, which accounts for nonidealities in coupling between the input and output ports.

\begin{figure}[ht]
  \begin{subfigure}[t]{0.45\textwidth}
    \centering
    \includegraphics[width=\textwidth]{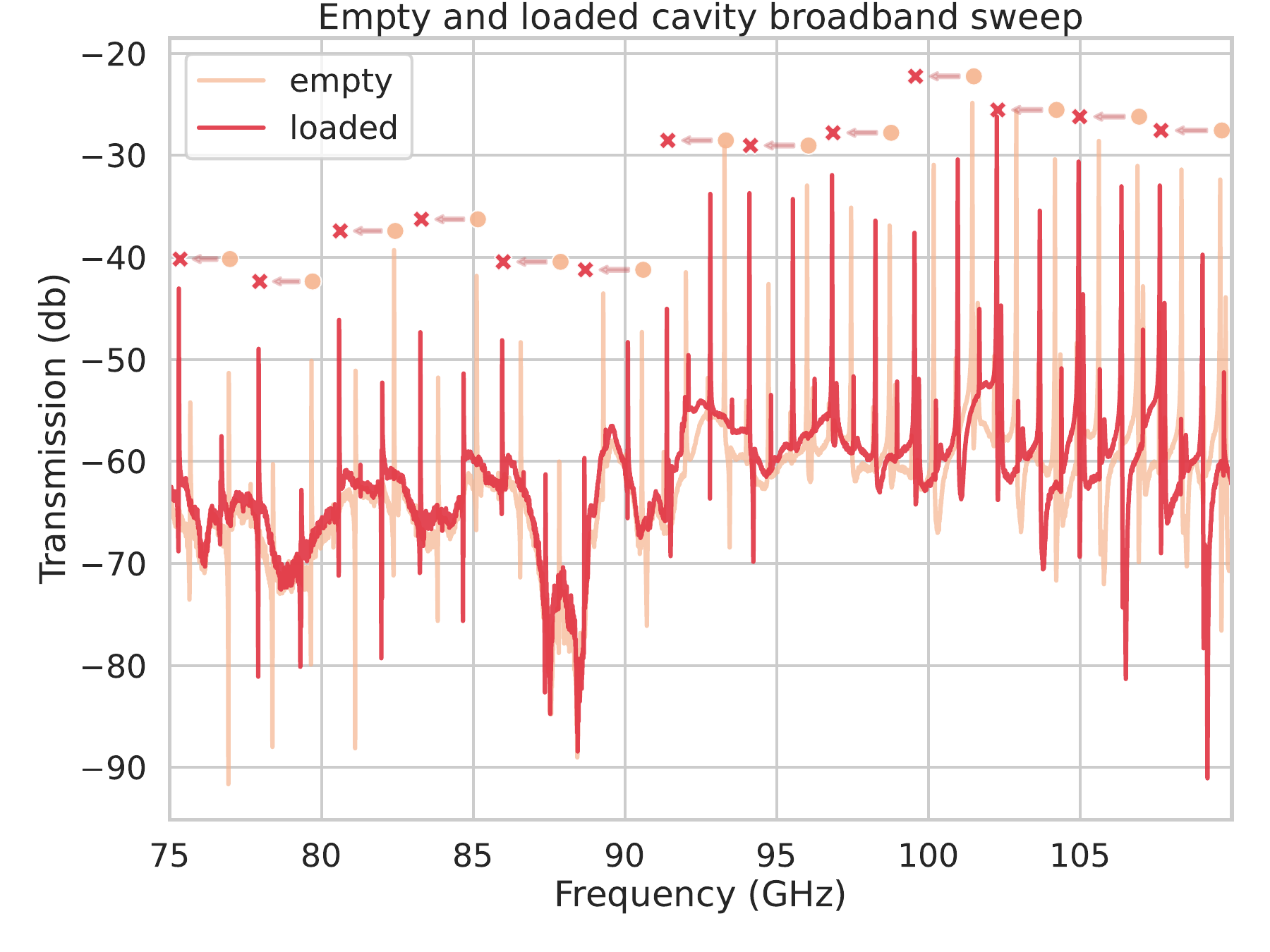}
    \caption{}
    \label{fig:broadband}
  \end{subfigure}
  \begin{subfigure}[t]{.45\textwidth}
    \centering
    \includegraphics[width=\linewidth]{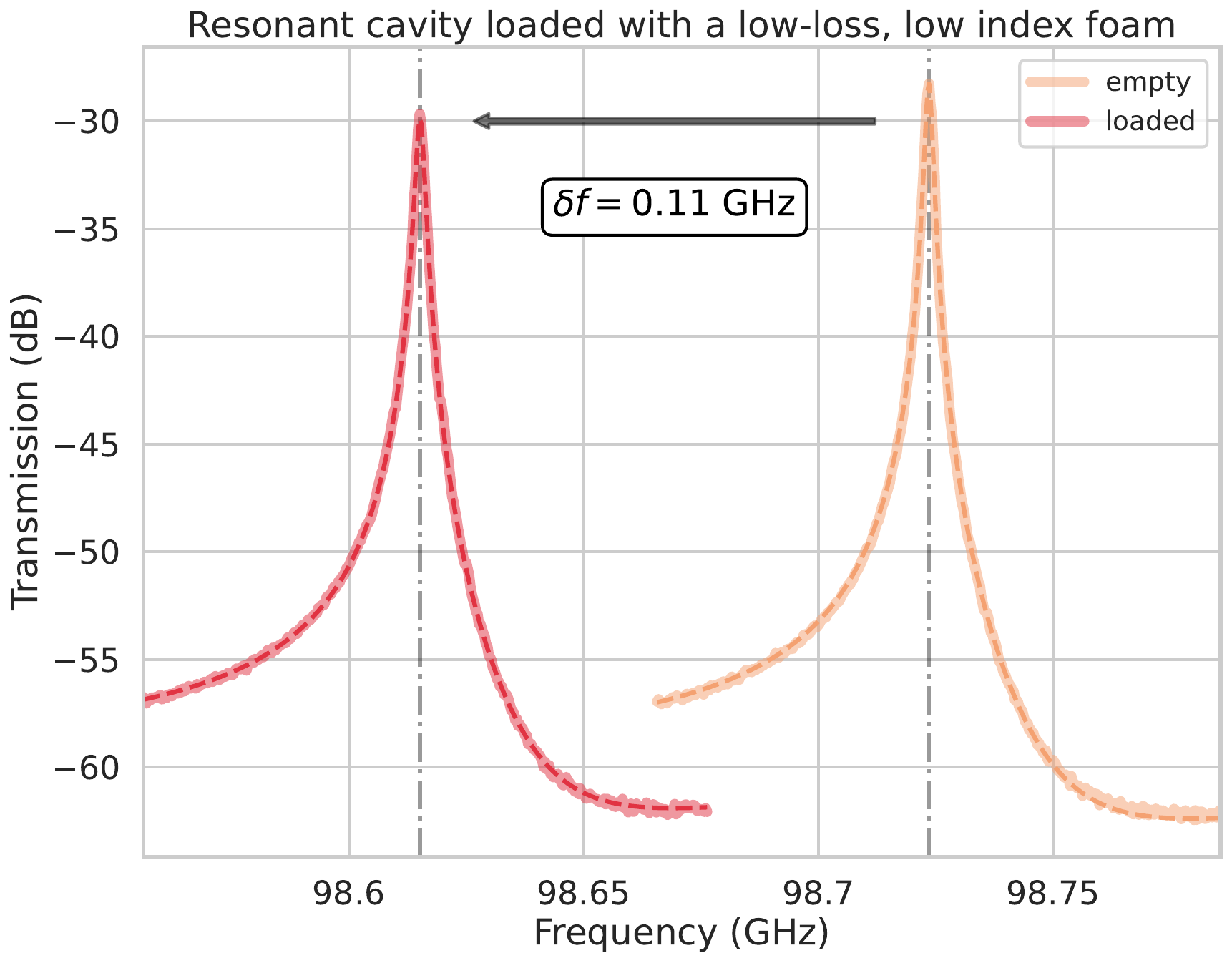}
    \caption{}
    \label{fig:shift}
  \end{subfigure}

  \caption{(\subref{fig:broadband}) Example broadband sweep of empty and loaded cavity, demonstrating the shift in resonance modes due to increasing cavity electrical length. (\subref{fig:shift}) Example resonant mode measurement (solid lines) and fit (dashed lines). The cavity was loaded with a low-loss, low index material such that the shift in frequency was small and the decrease in $Q$ nearly zero. The background is dominated by the leakage indicated in Fig. \ref{fig:schematic}.}
  \label{fig:cavity_sweeps}
\end{figure}

As seen in Eqn. \eqref{eq:losstan}, computing loss tangent requires carefully measuring the fundamental modes of the dielectric-loaded and empty cavity. In particular, we must measure their resonance frequencies (a measure of cavity length, including sample electrical length) and their quality factors (a measure of the energy stored in a given mode) \cite{yuMeasurementPermittivityMeans1982,shuMillimeterWaveMeasurement2015}.

We fix cavity length and sweep the band while recording the cavity response. Fig. \ref{fig:broadband} gives an example of the typical cavity response. First, we must find the cavity length and mirror radius of curvature. Both parameters are obtained by first measuring each fundamental mode of the empty cavity. Practically, we compute the fundamental mode resonance frequencies by iteratively solving Eqn. \eqref{eq:empty_res} using priors on $D$ and $R$ from the cavity design. Disambiguation of the axial mode numbers of the resonances requires moderate precision on the priors $D$ and $R$, which is easily exceeded by the manufacturing tolerances. For each measured fundamental mode, we fit a Fano lineshape to the cavity response and infer the mode frequencies and quality factors. From these direct measurements of mode frequencies, we can accurately deduce the cavity length and mirror radius of curvature.

With the cavity length dimensions in hand, we then load the cavity with a dielectric sample. After measuring each fundamental mode in the band of interest and inferring their center frequencies, $f_m$, and quality factors, $Q_L$, using the Fano model, we must identify them with the estimated resonances, $f_{L,i}$, as computed from the Eqn. \eqref{eqn:loaded_res}, using mechanical measurements of the thickness of the sample, and our priors on the refractive index of the material from previously published measurements\cite{lambMiscellaneousDataMaterials1996, goldsmithQuasiopticalSystemsGaussian1998}. As for the empty cavity, we only need sufficient precision in these measurements to disambiguate the axial mode number of each resonance. We then apply the interface and mirror corrections,

\begin{align}
  \label{eq:corrected_res}
  f = f_m - f_m (f_{\textrm{interface}}+f_{\textrm{mirror}}),
\end{align}
to arrive at the loaded cavity resonant frequencies.

\begin{figure}
  \centering
  \includegraphics[width=.65\linewidth]{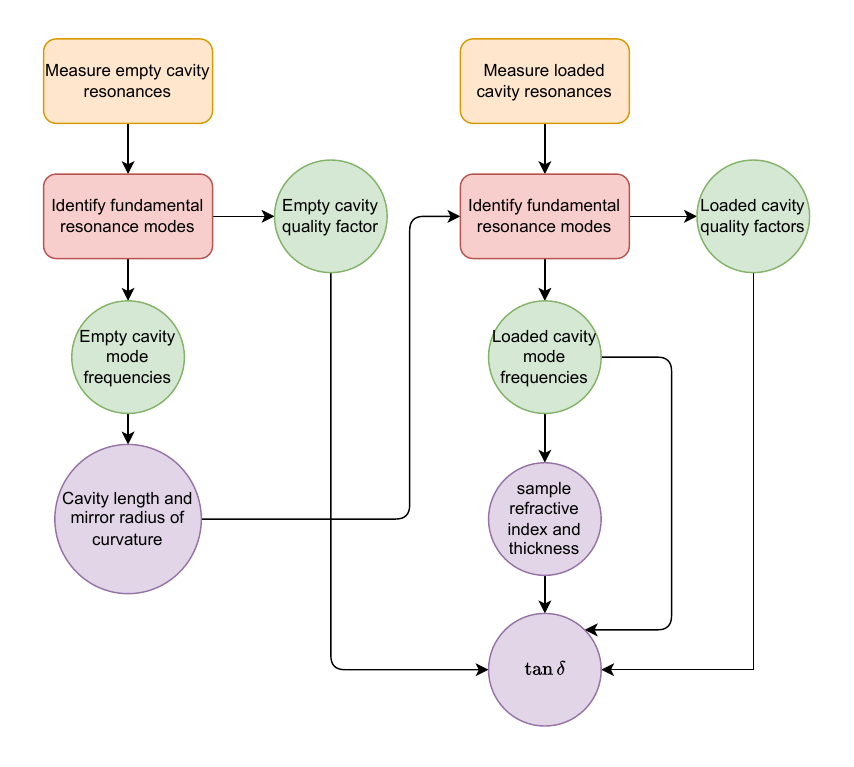}
  \caption{Flowchart illustrating the process to compute sample index and $\tan\delta$.}
  \label{fig:flowchart}
\end{figure}

Comparing the shift in resonance frequencies in the empty and loaded cavity, we infer the electrical length of the sample. Fig. \ref{fig:shift} is an example trace demonstrating the shift in resonance. Finally, the cavity parameters, sample parameters, and cavity empty and loaded fundamental mode frequencies and quality factors are used to compute material loss.

The measurement procedure is then repeated for samples of the same material, but of increasing thicknesses. Additionally, the accuracy of the resonant frequency and interface conditions depends on the accuracy of a direct measurement of the sample thickness. In this work, we machined $2''\times2''$ flat samples that were $\SI{2}{\mm}$, $\SI{3}{\mm}$, $\SI{5}{\mm}$, and $\SI{8}{\mm}$ thick, which were then measured with a micrometer to obtain a sample thickness for inferring sample index and loss. Fig. \ref{fig:flowchart} summarizes the above process for measuring sample index and loss tangent in the resonant cavity.

\subsection{Measurement systematics}
Given the reliance of this analysis on the flatness of the sample interface, thickness, and, to some extent, surface roughness, we performed tests on the repeatability of the quality factors of the fundamental modes.

In order to assess quality factor measurement repeatability, we devised a set of measurements of a single fundamental mode, wherein we perturbed the sample placement in some fashion. Repeatedly removing and placing the sample gives a sense of the repeatability of the cavity quality factor (and thus loss) due to the sample placement procedure. In this vein, we can change the sample placement procedure to asses the sensitivity of the cavity to other perturbations, such as sample rotation. In principle, this could be sensitive to sample birefringence. Further, by flipping the sample, we can see if the sample-cavity interface affects measurement repeatability. We might expect this effect to be stronger than that caused by rotation, as one can imagine the surface texture or curvature, which can vary due to machining artifacts, will impact the measurement. Finally, as a control, we took a large number of measurements while keeping the sample undisturbed.  The percent RMS of the repeated sample placement and rotated sample single-mode quality factor measurements were all consistent with that derived from the series of unperturbed sample quality factor measurements. Nonflat samples had a detectable difference in quality factor when changing which face interfaces with the cavity.

%% file: sec/results.tex
\section{Open resonator measurements of refractive index and dielectric loss}
\label{sec:results}

\begin{figure}[!ht]
    \centering
    \includegraphics[width=.95\linewidth]{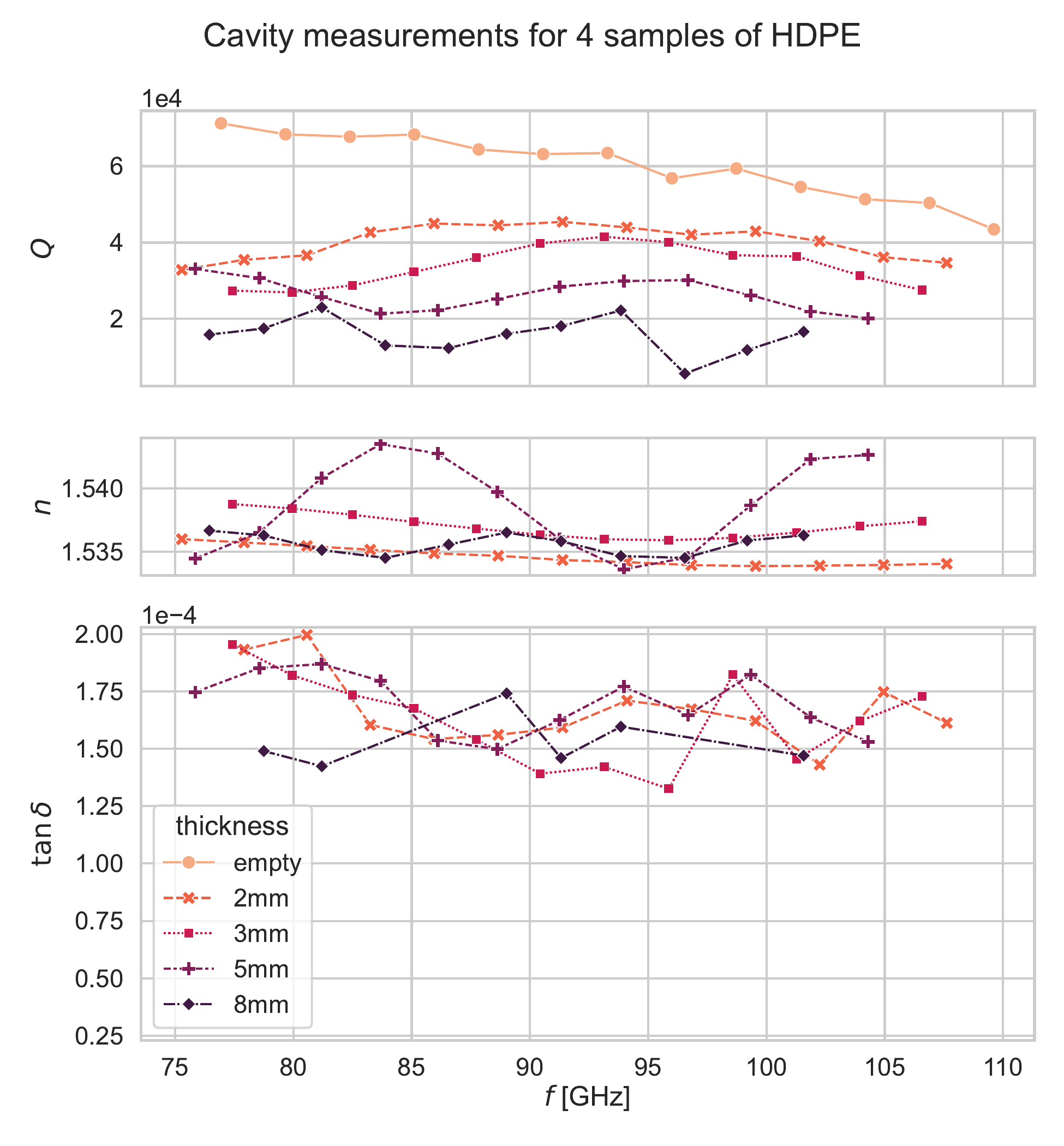}
    \caption{Measured cavity quality factor, sample index, and loss tangent for 4 samples of HDPE at room temperature.}
    \label{fig:HDPE_all_samples}
\end{figure}

After following the procedure described in the previous section, we obtain the index of refraction and loss tangent of several plastics in the WR-10 band (\SI{75}{\GHz}-\SI{110}{\GHz}) at room temperature. We differentiate between annealed and non-annealed HDPE to demonstrate the utility of this technique in exploring the effect of the thermal history on the complex permittivity of bulk plastics. Annealing is common when fabricating precision-machined optics like lenses from thermoplastics such as HDPE because the annealing process relieves internal stresses, increasing the dimensional stability of the optic, and can also reduce birefringence due to asymmetry in the material manufacturing processes. Due to the changes in density driven by internal stresses of the plastic, we may expect a change in the complex permittivity. 

Fig. \ref{fig:HDPE_all_samples} shows the per-mode index and loss measurements for the empty cavity and four samples of HDPE. Results reported in Table \ref{tab:results} are the mean and standard deviation taken across all measurements (per mode and thickness) for a given material. The spread in the loss tangent measurements is consistent with uncertainty in the sample thickness. The results for the refractive index of HDPE and UHMWPE are consistent with those reported in other works. Even after accounting for frequency scaling, the loss tangents reported here are somewhat lower than those reported in previous works, but are overall consistent, considering the nontrivial dependence of loss on the specifics of a polymer's manufacturing and thermal history \cite{shitvovBroadbandCoatedLens2022,dalessandroUltraHighMolecular2018a,lambMiscellaneousDataMaterials1996}.
\setlength{\tabcolsep}{10pt} 
\renewcommand{\arraystretch}{1.5} 
\begin{table}[!ht]
    \centering
    \begin{tabular}{l r r c}
        \toprule
        \textbf{material} & \textbf{$n$}                                  & \textbf{$\tan\delta$ \ $(\times 10^{-4})$} & \textbf{loss (\si{\decibel})} \\
        \midrule
        HDPE              & \num[round-precision=1]{1.536630 \pm 0.003}\  & \num[round-precision=2]{1.66 \pm .18}      & -28                           \\
        annealed HDPE     & \num[round-precision=1]{1.541203 \pm .004}    & \num[round-precision=2]{1.48 \pm .24}      & -29                           \\
        UHMWPE            & \num[round-precision=1]{1.525534 \pm 0.007}   & \num[round-precision=2]{1.43 \pm .21}      & -29                           \\
        LDPE              & \                                             &                                            & -49                           \\
        \bottomrule
    \end{tabular}
    \
    \caption{Indices and losses of several plastics at room temperature. Loss is reported as power lost from the cavity in a single round trip. A measurement of \SI{25}{\um} thick LDPE is included to illustrate the capability of the cavity to measure thin, low-loss materials.}
    \label{tab:results}
\end{table}

%% file: sec/conclusion.tex
\section{Conclusions \& Future Research}
\label{sec:conclusion}

We have presented a case for the metrological application of mm-wave resonant cavities to the measurement of dielectrics used to make mm-wave optics. We can robustly model the fundamental modes of a loaded and unloaded quasioptical resonant cavity, confirming that resonant cavities can be a useful tool in determining the index and loss of dielectrics. We have presented initial results for the indices and loss tangents of HDPE, annealed HDPE, and UHMWPE.

Aside from the dielectric parameters reported here, we want to understand room temperature losses of various commonly used and potential candidate materials for CMB telescope components. We are acutely interested in the behavior of dielectrics at cryogenic temperatures due to their use in mm-wave experiments such as CMB polarimeters. Continuing work on this apparatus involves placing the cavity in a quick-turnaround cryostat in order to measure index and losses from \SI{300}{\kelvin} down to \SI{4}{\kelvin}.

Further work must be done to model the higher-order modes of the loaded cavity. Higher-order modes should have a slightly different dependence on index and sample thickness. Due to this difference, higher-order modes can potentially provide further constraints on sample thickness. As sample thickness is the dominant systematic in the reported index and loss measurements, improvements in the understanding of the higher-order modes will potentially decrease uncertainty.

While we can measure thin, low-loss materials down to single-trip losses of $\SI{-49}{\dB}$, to explore even lower losses we must increase the cavity quality factor. As discussed, the natural path to increase cavity quality factor is by decreasing aperture coupling. However, since our background is dominated by direct transmission, increases in quality factor are commensurate to increases in the background. To complicate matters, this background also depends on the seating pressure of the coupling block; the choke between the it and the spherical mirror relies on pressure. As a result, the background levels are sensitive to temperature fluctuations. Future improvements on cavity design will involve reducing this leakage, possibly by utilizing a photonic crystalline structure to introduce a photonic bandgap that acts to suppress leakage through the waveguide-mirror interface and direct leakage across the split-block waveguide\cite{Hesler2001}.

%% file: main.bbl
\begin{thebibliography}{10}

\bibitem{johnsonKeyScienceGoals2023}
Johnson, M.~D., Akiyama, K., Blackburn, L., Bouman, K.~L., Broderick, A.~E., Cardoso, V., Fender, R.~P., Fromm, C.~M., Galison, P., G{\'o}mez, J.~L., Haggard, D., Lister, M.~L., Lobanov, A.~P., Markoff, S., Narayan, R., Natarajan, P., Nichols, T., Pesce, D.~W., Younsi, Z., Chael, A., Chatterjee, K., Chaves, R., Doboszewski, J., Dodson, R., Doeleman, S.~S., Elder, J., Fitzpatrick, G., Haworth, K., Houston, J., Issaoun, S., Kovalev, Y.~Y., Levis, A., Lico, R., Marcoci, A., Martens, N. C.~M., Nagar, N.~M., Oppenheimer, A., Palumbo, D. C.~M., Ricarte, A., Rioja, M.~J., Roelofs, F., Thresher, A.~C., Tiede, P., Weintroub, J., and Wielgus, M., ``Key {{Science Goals}} for the {{Next-Generation Event Horizon Telescope}},'' {\em Galaxies}~{\bf 11},  61 (June 2023).

\bibitem{karkareSPTSLIMLineIntensity2022}
Karkare, K.~S., Anderson, A.~J., Barry, P.~S., Benson, B.~A., Carlstrom, J.~E., Cecil, T., Chang, C.~L., Dobbs, M.~A., Hollister, M., Keating, G.~K., Marrone, D.~P., McMahon, J., Montgomery, J., Pan, Z., Robson, G., Rouble, M., Shirokoff, E., and Smecher, G., ``{{SPT-SLIM}}: {{A Line Intensity Mapping Pathfinder}} for the {{South Pole Telescope}},'' {\em Journal of Low Temperature Physics}~{\bf 209},  758--765 (Dec. 2022).

\bibitem{bicep/keckcollaborationImprovedConstraintsPrimordial2021}
{BICEP/Keck Collaboration}, Ade, P. A.~R., Ahmed, Z., Amiri, M., Barkats, D., Thakur, R.~B., Bischoff, C.~A., Beck, D., Bock, J.~J., Boenish, H., Bullock, E., Buza, V., Cheshire, J.~R., Connors, J., Cornelison, J., Crumrine, M., Cukierman, A., Denison, E.~V., Dierickx, M., Duband, L., Eiben, M., Fatigoni, S., Filippini, J.~P., Fliescher, S., {Goeckner-Wald}, N., Goldfinger, D.~C., Grayson, J., Grimes, P., Hall, G., Halal, G., Halpern, M., Hand, E., Harrison, S., Henderson, S., Hildebrandt, S.~R., Hilton, G.~C., Hubmayr, J., Hui, H., Irwin, K.~D., Kang, J., Karkare, K.~S., Karpel, E., Kefeli, S., Kernasovskiy, S.~A., Kovac, J.~M., Kuo, C.~L., Lau, K., Leitch, E.~M., Lennox, A., Megerian, K.~G., Minutolo, L., Moncelsi, L., Nakato, Y., Namikawa, T., Nguyen, H.~T., O'Brient, R., Ogburn, R.~W., Palladino, S., Prouve, T., Pryke, C., Racine, B., Reintsema, C.~D., Richter, S., Schillaci, A., Schwarz, R., Schmitt, B.~L., Sheehy, C.~D., Soliman, A., Germaine, T.~{\relax St}., Steinbach, B., Sudiwala, R.~V., Teply, G.~P., Thompson, K.~L., Tolan, J.~E., Tucker, C., Turner, A.~D., Umilt{\`a}, C., Verg{\`e}s, C., Vieregg, A.~G., Wandui, A., Weber, A.~C., Wiebe, D.~V., Willmert, J., Wong, C.~L., Wu, W. L.~K., Yang, H., Yoon, K.~W., Young, E., Yu, C., Zeng, L., Zhang, C., and Zhang, S., ``Improved {{Constraints}} on {{Primordial Gravitational Waves}} using {{Planck}}, {{WMAP}}, and {{BICEP}}/{{Keck Observations}} through the 2018 {{Observing Season}},'' {\em Physical Review Letters}~{\bf 127},  151301 (Oct. 2021).

\bibitem{choiRadiotransparentMultilayerInsulation2013}
Choi, J., Ishitsuka, H., Mima, S., Oguri, S., Takahashi, K., and Tajima, O., ``Radio-transparent multi-layer insulation for radiowave receivers,'' {\em Review of Scientific Instruments}~{\bf 84},  114502 (Nov. 2013).

\bibitem{dierickxPlasticLaminateAntireflective2023}
Dierickx, M., Ade, P. A.~R., Ahmed, Z., Amiri, M., Barkats, D., Basu~Thakur, R., Bischoff, C.~A., Beck, D., Bock, J.~J., Buza, V., Cheshire, J., Connors, J., Cornelison, J., Crumrine, M., Cukierman, A., Denison, E., Duband, L., Eiben, M., Fatigoni, S., Filippini, J.~P., {Goeckner-Wald}, N., Goldfinger, D.~C., Grayson, J.~A., Grimes, P., Hall, G., Halal, G., Halpern, M., Hand, E., Harrison, S., Henderson, S., Hildebrandt, S.~R., Hilton, G.~C., Hubmayr, J., Hui, H., Irwin, K.~D., Kang, J., Karkare, K.~S., Kefeli, S., Kovac, J.~M., Kuo, C.~L., Lau, K., Leitch, E.~M., Lennox, A., Megerian, K.~G., Minutolo, L., Moncelsi, L., Nakato, Y., Namikawa, T., Nguyen, H.~T., O'Brient, R., Palladino, S., Petroff, M., Precup, N., Prouve, T., Pryke, C., Racine, B., Reintsema, C.~D., Santalucia, D., Schillaci, A., Schmitt, B.~L., Singari, B., Soliman, A., Germaine, T.~{\relax St}., Steinbach, B., Sudiwala, R.~V., Thompson, K.~L., Tucker, C., Turner, A.~D., Umilt{\`a}, C., Verges, C., Vieregg, A.~G., Wandui, A., Weber, A.~C., Wiebe, D.~V., Willmert, J., Wu, W. L.~K., Yang, E., Yoon, K.~W., Young, E., Yu, C., Zeng, L., Zhang, C., and Zhang, S., ``Plastic {{Laminate Antireflective Coatings}} for {{Millimeter-Wave Optics}} in {{BICEP Array}},'' {\em Journal of Low Temperature Physics}~{\bf 211},  366--374 (June 2023).

\bibitem{barkatsUltrathinLargeapertureVacuum2018}
Barkats, D., Dierickx, M.~I., Kovac, J.~M., Pentacoff, C., Ade, P. a.~R., Ahmed, Z., Aikin, R.~W., Alexander, K.~D., Benton, S.~J., Bischoff, C.~A., Bock, J.~J., {Bowens-Rubin}, R., Brevik, J.~A., Buder, I., Bullock, E., Buza, V., Connors, J., Cornelison, J., Crill, B.~P., Crumrine, M., Duband, L., Dvorkin, C., Filippini, J.~P., Fliescher, S., Grayson, J.~A., Hall, G., Halpern, M., Harrison, S.~A., Hildebrandt, S.~R., Hilton, G.~C., Hui, H., Irwin, K.~D., Kang, J., Karkare, K.~S., Karpel, E., Kaufman, J.~P., Keating, B.~G., Kefeli, S., Kernasovskiy, S.~A., Kuo, C.~L., Lau, K., Larsen, N.~A., Leitch, E.~M., Lueker, M., Megerian, K.~G., Moncelsi, L., Namikawa, T., Nguyen, H.~T., O'Brient, R., Iv, R. W.~O., Palladino, S., Pryke, C., Racine, B., Richter, S., Schwarz, R., Schillaci, A., Sheehy, C.~D., Soliman, A., Germaine, T.~S., Staniszewski, Z.~K., Steinbach, B., Sudiwala, R.~V., Teply, G.~P., Thompson, K.~L., Tolan, J.~E., Tucker, C., Turner, A.~D., Umilt{\`a}, C., Vieregg, A.~G., Wandui, A., Weber, A.~C., Wiebe, D.~V., Willmert, J., Wong, C.~L., Wu, W. L.~K., Yang, H., Yoon, K.~W., and Zhang, C., ``Ultra-thin large-aperture vacuum windows for millimeter wavelengths receivers,'' in [{\em Millimeter, {{Submillimeter}}, and {{Far-Infrared Detectors}} and {{Instrumentation}} for {{Astronomy IX}}}{\nolinebreak\hspace{0.1em}]},   {\bf 10708},  587--600, SPIE (July 2018).

\bibitem{carterLowlossSiliconOptical2024}
Carter, K.~J., Tong, C.~E., Zeng, L., Grimes, P., and Kimberk, R., ``A low-loss silicon optical diplexer for millimeter and submillimeter radio astronomy,'' in [{\em Millimeter, {{Submillimeter}}, and {{Far-Infrared Detectors}} and {{Instrumentation}} for {{Astronomy XII}}}{\nolinebreak\hspace{0.1em}]},   {\bf 13102},  716--719, SPIE (Aug. 2024).

\bibitem{sobrinDesignIntegratedPerformance2022}
Sobrin, J.~A., Anderson, A.~J., Bender, A.~N., Benson, B.~A., Dutcher, D., Foster, A., {Goeckner-Wald}, N., Montgomery, J., Nadolski, A., Rahlin, A., Ade, P. A.~R., Ahmed, Z., Anderes, E., Archipley, M., Austermann, J.~E., Avva, J.~S., Aylor, K., Balkenhol, L., Barry, P.~S., Thakur, R.~B., Benabed, K., Bianchini, F., Bleem, L.~E., Bouchet, F.~R., Bryant, L., Byrum, K., Carlstrom, J.~E., Carter, F.~W., Cecil, T.~W., Chang, C.~L., Chaubal, P., Chen, G., Cho, H.-M., Chou, T.-L., Cliche, J.-F., Crawford, T.~M., Cukierman, A., Daley, C., de~Haan, T., Denison, E.~V., Dibert, K., Ding, J., Dobbs, M.~A., Everett, W., Feng, C., Ferguson, K.~R., Fu, J., Galli, S., Gambrel, A.~E., Gardner, R.~W., Gualtieri, R., Guns, S., Gupta, N., Guyser, R., Halverson, N.~W., {Harke-Hosemann}, A.~H., Harrington, N.~L., Henning, J.~W., Hilton, G.~C., Hivon, E., Holder, G.~P., Holzapfel, W.~L., Hood, J.~C., Howe, D., Huang, N., Irwin, K.~D., Jeong, O.~B., Jonas, M., Jones, A., Khaire, T.~S., Knox, L., Kofman, A.~M., Korman, M., Kubik, D.~L., Kuhlmann, S., Kuo, C.-L., Lee, A.~T., Leitch, E.~M., Lowitz, A.~E., Lu, C., Meyer, S.~S., Michalik, D., Millea, M., Natoli, T., Nguyen, H., Noble, G.~I., Novosad, V., Omori, Y., Padin, S., Pan, Z., Paschos, P., Pearson, J., Posada, C.~M., Prabhu, K., Quan, W., Reichardt, C.~L., Riebel, D., Riedel, B., Rouble, M., Ruhl, J.~E., Saliwanchik, B., Sayre, J.~T., Schiappucci, E., Shirokoff, E., Smecher, G., Stark, A.~A., Stephen, J., Story, K.~T., Suzuki, A., Tandoi, C., Thompson, K.~L., Thorne, B., Tucker, C., Umilta, C., Vale, L.~R., Vanderlinde, K., Vieira, J.~D., Wang, G., Whitehorn, N., Wu, W. L.~K., Yefremenko, V., Yoon, K.~W., and Young, M.~R., ``The {{Design}} and {{Integrated Performance}} of {{SPT-3G}},'' {\em The Astrophysical Journal Supplement Series}~{\bf 258},  42 (Feb. 2022).

\bibitem{adeBicepKeckXV2022}
Ade, P. A.~R., Ahmed, Z., Amiri, M., Barkats, D., Thakur, R.~B., Bischoff, C.~A., Beck, D., Bock, J.~J., Boenish, H., Bullock, E., Buza, V., IV, J. R.~C., Connors, J., Cornelison, J., Crumrine, M., Cukierman, A., Denison, E.~V., Dierickx, M., Duband, L., Eiben, M., Fatigoni, S., Filippini, J.~P., Fliescher, S., {Goeckner-Wald}, N., Goldfinger, D.~C., Grayson, J., Grimes, P., Hall, G., Halal, G., Halpern, M., Hand, E., Harrison, S., Henderson, S., Hildebrandt, S.~R., Hilton, G.~C., Hubmayr, J., Hui, H., Irwin, K.~D., Kang, J., Karkare, K.~S., Karpel, E., Kefeli, S., Kernasovskiy, S.~A., Kovac, J.~M., Kuo, C.~L., Lau, K., Leitch, E.~M., Lennox, A., Megerian, K.~G., Minutolo, L., Moncelsi, L., Nakato, Y., Namikawa, T., Nguyen, H.~T., O'Brient, R., IV, R. W.~O., Palladino, S., Prouve, T., Pryke, C., Racine, B., Reintsema, C.~D., Richter, S., Schillaci, A., Schwarz, R., Schmitt, B.~L., Sheehy, C.~D., Soliman, A., Germaine, T.~S., Steinbach, B., Sudiwala, R.~V., Teply, G.~P., Thompson, K.~L., Tolan, J.~E., Tucker, C., Turner, A.~D., Umilt{\`a}, C., Verg{\`e}s, C., Vieregg, A.~G., Wandui, A., Weber, A.~C., Wiebe, D.~V., Willmert, J., Wong, C.~L., Wu, W. L.~K., Yang, H., Yoon, K.~W., Young, E., Yu, C., Zeng, L., Zhang, C., Zhang, S., and Collaboration), B.~K., ``Bicep/{{Keck XV}}: {{The Bicep3 Cosmic Microwave Background Polarimeter}} and the {{First Three-year Data Set}},'' {\em The Astrophysical Journal}~{\bf 927},  77 (Mar. 2022).

\bibitem{paineProcessingCalibrationSubmillimeter2013}
Paine, S.~N. and Turner, D.~D., ``Processing and {{Calibration}} of {{Submillimeter Fourier Transform Radiometer Spectra From}} the {{RHUBC-II Campaign}},'' {\em IEEE Transactions on Geoscience and Remote Sensing}~{\bf 51},  5187--5198 (Dec. 2013).

\bibitem{weirAutomaticMeasurementComplex1974}
Weir, W., ``Automatic measurement of complex dielectric constant and permeability at microwave frequencies,'' {\em Proceedings of the IEEE}~{\bf 62},  33--36 (Jan. 1974).

\bibitem{baker-jarvisImprovedTechniqueDetermining1990}
{Baker-Jarvis}, J., Vanzura, E., and Kissick, W., ``Improved technique for determining complex permittivity with the transmission/reflection method,'' {\em IEEE Transactions on Microwave Theory and Techniques}~{\bf 38},  1096--1103 (Aug. 1990).

\bibitem{chamberlainDeterminationRefractiveIndex1969}
Chamberlain, J., Gibbs, J.~E., and Gebbie, H.~A., ``The determination of refractive index spectra by fourier spectrometry,'' {\em Infrared Physics}~{\bf 9},  185--209 (Dec. 1969).

\bibitem{nadolskiBroadbandMillimeterwaveAntireflection2020}
Nadolski, A., Vieira, J.~D., Sobrin, J.~A., Kofman, A.~M., Ade, P. a.~R., Ahmed, Z., Anderson, A.~J., Avva, J.~S., Thakur, R.~B., Bender, A.~N., Benson, B.~A., Bryant, L., Carlstrom, J.~E., Carter, F.~W., Cecil, T.~W., Chang, C.~L., Cheshire, J.~R., Chesmore, G.~E., Cliche, J.~F., Cukierman, A., de~Haan, T., Dierickx, M., Ding, J., Dutcher, D., Everett, W., Farwick, J., Ferguson, K.~R., Florez, L., Foster, A., Fu, J., Gallicchio, J., Gambrel, A.~E., Gardner, R.~W., Groh, J.~C., Guns, S., Guyser, R., Halverson, N.~W., {Harke-Hosemann}, A.~H., Harrington, N.~L., Harris, R.~J., Henning, J.~W., Holzapfel, W.~L., Howe, D., Huang, N., Irwin, K.~D., Jeong, O., Jonas, M., Jones, A., Korman, M., Kovac, J., Kubik, D.~L., Kuhlmann, S., Kuo, C.-L., Lee, A.~T., Lowitz, A.~E., McMahon, J., Meier, J., Meyer, S.~S., Michalik, D., Montgomery, J., Natoli, T., Nguyen, H., Noble, G.~I., Novosad, V., Padin, S., Pan, Z., Paschos, P., Pearson, J., Posada, C.~M., Quan, W., Rahlin, A., Riebel, D., Ruhl, J.~E., Sayre, J.~T., Shirokoff, E., Smecher, G., Stark, A.~A., Stephen, J., Story, K.~T., Suzuki, A., Tandoi, C., Thompson, K.~L., Tucker, C., Vanderlinde, K., Wang, G., Whitehorn, N., Yefremenko, V., Yoon, K.~W., and Young, M.~R., ``Broadband, millimeter-wave antireflection coatings for large-format, cryogenic aluminum oxide optics,'' {\em Applied Optics}~{\bf 59},  3285--3295 (Apr. 2020).

\bibitem{cullenAccurateMeasurementPermittivity1997}
Cullen, A.~L., Yu, P.~K., and Barlow, H. E.~M., ``The accurate measurement of permittivity by means of an open resonator,'' {\em Proceedings of the Royal Society of London. A. Mathematical and Physical Sciences}~{\bf 325},  493--509 (Jan. 1997).

\bibitem{yuMeasurementPermittivityMeans1982}
Yu, P.~K. and Cullen, A.~L., ``Measurement of permittivity by means of an open resonator. {{I}}. {{Theoretical}},'' {\em Proceedings of the Royal Society of London. A. Mathematical and Physical Sciences}~{\bf 380},  49--71 (Mar. 1982).

\bibitem{hirvonenMeasurementDielectrics1001996}
Hirvonen, T., Vainikainen, P., Lozowski, A., and Raisanen, A., ``Measurement of dielectrics at 100 {{GHz}} with an open resonator connected to a network analyzer,'' {\em IEEE Transactions on Instrumentation and Measurement}~{\bf 45},  780--786 (Aug. 1996).

\bibitem{petersanMeasurementResonantFrequency1998}
Petersan, P.~J. and Anlage, S.~M., ``Measurement of {{Resonant Frequency}} and {{Quality Factor}} of {{Microwave Resonators}}: {{Comparison}} of {{Methods}},'' {\em Journal of Applied Physics}~{\bf 84},  3392--3402 (Sept. 1998).

\bibitem{shuMillimeterWaveMeasurement2015}
Shu, G.~X., Luo, Y., Zhang, Q.~S., Su, J., Wang, L., Xu, Y., and Wang, S.~F., ``Millimeter {{Wave Measurement}} of the {{Low-Loss Dielectric}} in {{Vacuum Electronic Devices}} with {{Reflection-Type Hemispherical Open Resonator}},'' {\em Journal of Infrared, Millimeter, and Terahertz Waves}~{\bf 36},  556--568 (June 2015).

\bibitem{lambMiscellaneousDataMaterials1996}
Lamb, J., ``Miscellaneous data on materials for millimetre and submillimetre optics,'' {\em International Journal of Infrared and Millimeter Waves}~{\bf 17},  1997--2034 (Dec. 1996).

\bibitem{goldsmithQuasiopticalSystemsGaussian1998}
Goldsmith, P.~F.,  [{\em Quasioptical {{Systems}}: {{Gaussian Beam Quasioptical Propogation}} and {{Applications}}}{\nolinebreak\hspace{0.1em}]}, Wiley-IEEE Press, Piscataway, NJ, 1st edition~ed. (Jan. 1998).

\bibitem{shitvovBroadbandCoatedLens2022}
Shitvov, A., Savini, G., Hargrave, P.~C., Ade, P. A.~R., Tucker, C.~E., Sudiwala, R.~V., Zhang, J., Gudmundsson, J.~E., Winter, B., Pisano, G., and van~der Vorst, M., ``Broadband coated lens solutions for {{FIR-mm-wave}} instruments,'' in [{\em Millimeter, {{Submillimeter}}, and {{Far-Infrared Detectors}} and {{Instrumentation}} for {{Astronomy XI}}}{\nolinebreak\hspace{0.1em}]},   {\bf 12190},  203--224, SPIE (Aug. 2022).

\bibitem{dalessandroUltraHighMolecular2018a}
D'Alessandro, G., Paiella, A., Coppolecchia, A., Castellano, M.~G., Colantoni, I., {de Bernardis}, P., Lamagna, L., and Masi, S., ``Ultra high molecular weight polyethylene: {{Optical}} features at millimeter wavelengths,'' {\em Infrared Physics and Technology}~{\bf 90},  59--65 (May 2018).

\bibitem{Hesler2001}
Hesler, J., ``A photonic crystal joint ({PCJ}) for metal waveguides,'' in [{\em 2001 IEEE MTT-S International Microwave Sympsoium Digest (Cat. No.01CH37157)}{\nolinebreak\hspace{0.1em}]},   {\bf 2},  783--786 vol.2 (2001).

\end{thebibliography}
